\documentclass{iau_FM}
\usepackage{graphicx}

\title[Late-type stellar density profile] 
{The late-type stellar density profile in the Galactic Center: A statistical approach}

\author[S. N. Chappell \& A. M. Ghez]   
{S. N. Chappell$^1$, A. M. Ghez$^1$, T. Do$^1$, G. D. Martinez$^1$, S. Yelda$^1$, B. N. Sitarski$^1$, J. R. Lu$^2$, \and M. R. Morris$^1$}

\affiliation{$^1$UCLA, Department of Physics and Astronomy, Los Angeles, CA, 90095, USA \\[\affilskip]
$^2$Department of Astronomy, University of California, Berkeley Berkeley, CA, 94720, USA}

\pubyear{2016}
\jname{Astronomy in Focus, Volume 1} 
\editors{Steve Longmore, Geoff Bicknell, and Roland Crocker}
\begin{document}

\maketitle

\begin{abstract}
The late-type stellar population in the Galactic Center was first predicted to reside in a dynamically relaxed cusp (power law slope ranging from 3/2 to 7/4). However, observations - which rely on models to correct for projection effects - have suggested a flat distribution instead. The need for this correction is due to the lack of information regarding the line-of-sight distances. With a two decade long baseline in astrometric measurements, we are now able to measure significant projected radial accelerations, six of which are newly reported here, that directly constrain line-of-sight distances. Here we present a statistical approach to take advantage of this information and more accurately constrain the shape of the radial density profile of the late-type stellar population in the Galactic Center.

\end{abstract}

\firstsection 
\section{Summary}
 
The late-type stellar population ($\sim$1 Gyr) was initially theorized to lie in a dynamically relaxed cusp, with $\gamma$, the negative power law index, valued between $\frac{7}{4}$ and $\frac{3}{2}$ (\cite[Bachall \& Wolf 1976, 1977]{1977ApJ...216..883B}). Observations since have shown a density profile that is flat, or $\gamma$ $\sim$ 0 (\cite[Buchholz et al. 2009]{2009A&A...499..483B}; \cite[Do et al. 2009, 2013a]{2013ApJ...764..154D}; \cite[Bartko et al. 2010]{2010ApJ...708..834B}). However these previous observations were limited due to a lack of information regarding the line-of-sight distances.

With a two decade long baseline in astrometric measurements, we are now able to directly constrain stellar line-of-sight distances through measuring projected radial accelerations,
\begin{equation}
\mid z \mid = \bigg[\bigg(\frac{GMR}{a_R}\bigg)^{2/3} - R^2\bigg]^{1/2}.
\end{equation}
where $a_R$ is the projected radial acceleration and $R$ is the projected distance. Projected radial accelerations are measured by fitting polynomials to our astrometric observations, assuming a Keplerian potential defined solely by Sgr A* (M = 4.1 $\times$ $10^6$ $M_{\odot}$ and $r_{GC}$ = 8 kpc, \cite[Ghez et al. 2008]{2008ApJ...689.1044G}).

\begin{figure}[b]
\begin{center}
 \includegraphics[width=0.99\textwidth]{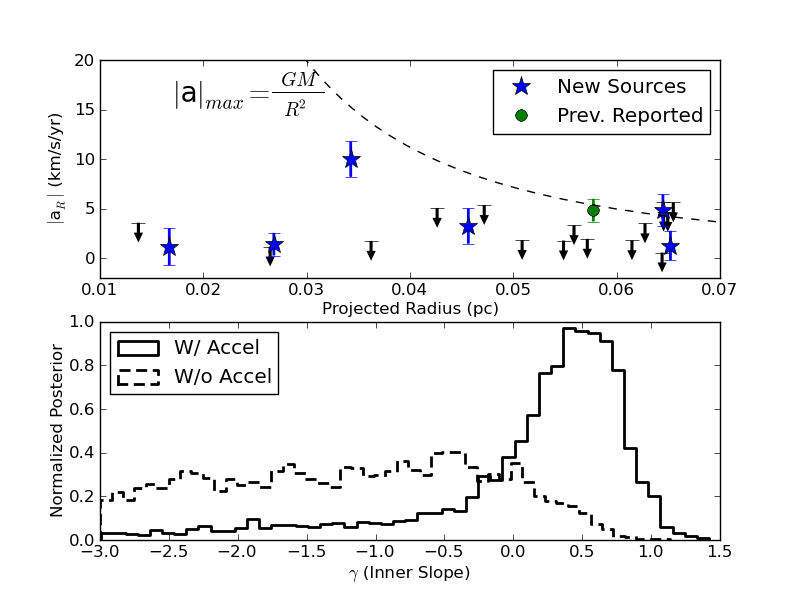} 
 \caption{(Top) Measured radial accelerations vs projected distance. New and previously reported significantly accelerating sources designated by stars and circles respectively. Three sigma upper limits are shown with arrows for the sources with non-significant acceleration measurements, all of which are newly reported here. The dashed line shows radial accelerations when line-of-sight distance is zero. (Bottom) Normalized posterior for $\gamma$. The solid line shows the posterior from the statistical approach with acceleration information, while the dashed line shows with just projected positions (no acceleration information). Unlike the other posterior, the inclusion of acceleration information shows a clear peak at $\gamma$ $\sim$ +0.43. For further discussion, see Chappell et al. 2017 (in prep.).}
   \label{both_fig}
\end{center}
\end{figure}

Seven of our 23 sources having $R$ $\textless$ 1.7 arcsec are found to have significant radial accelerations, and thus significant 3D distances, six of which are newly reported here (figure \ref{both_fig}, top). As not all of our sources have significant 3D distances, in order to take advantage of all the available information, we employ a statistical approach to fit a broken power law density profile to our sample of late-type stars,
\begin{equation}
\rho \propto \bigg( \frac{r}{r_{break}} \bigg)^{-\gamma} (1 + (r/r_{break})^{\delta})^{(\gamma - \alpha)/\delta} 
\end{equation}
where $\gamma$, $\alpha$, $\delta$, and $r_{break}$ are the inner slope, outer slope, sharpness of the transition, and the break radius respectively. The sample includes the GCOWS fields within 5" (\cite[Do et al. 2013a]{2013ApJ...764..154D}) and the projected positions between 5" and 15" from \cite[Sch{\"o}del et al. (2010)]{2010A&A...511A..18S}, to help constrain the outer parameters ($\alpha$ and $r_{break}$). We use Bayesian inference to fit for these parameters, with the likelihood ($\mathcal{L}$) being set to
the probability of a star having a given projected radial acceleration ($a_{R,i}$), times the probability of a star being at a given location, marginalized over the line-of-sight distance (z), summed over our sample, and each star weighted by its probability of being late-type ($p_{old}$, \cite[Do et al. 2013a]{2013ApJ...764..154D}). Flat priors are used for all radial density profile parameters.
\begin{equation}
\mathcal{L} = \prod^{stars}_{i} \bigg( \int p(a_{R,i} | x_i, y_i, z, M_{BH}, R_0) p(x_i, y_i, z | \gamma, \alpha, \delta, r_{break}, M_{BH}, R_0) dz\bigg)^{p_{old}}
\end{equation}

The resulting posterior for $\gamma$ is shown in figure \ref{both_fig} (bottom), see Chappell et al. 2017 (in prep.) for more details. The inclusion of acceleration information results in a clear peak, a $\gamma$ value of +0.43 $\pm$ 0.42, which suggests a cusp in the late-type stellar population that is shallower than initially theoretically predicted and steeper than previously suggested in observations. 

\begin{thebibliography}{}

\bibitem[Bahcall \& Wolf(1976)]{1976ApJ...209..214B} Bahcall, J.~N., \& Wolf, R.~A.\ 1976, \textit{ApJ}, 209, 214

\bibitem[Bahcall \& Wolf(1977)]{1977ApJ...216..883B} Bahcall, J.~N., \& Wolf, R.~A.\ 1977, \textit{ApJ}, 216, 883

\bibitem[Bartko et al.(2010)]{2010ApJ...708..834B} Bartko, H., Martins, F., Trippe, S., et al.\ 2010, \textit{ApJ}, 708, 834 

\bibitem[Buchholz et al.(2009)]{2009A&A...499..483B} Buchholz, R.~M., Sch{\"o}del, R., \& Eckart, A.\ 2009, \textit{A\&A}, 499, 483 

\bibitem[Do et al.(2009)]{2009ApJ...703.1323D} Do, T., Ghez, A.~M., Morris, M.~R., et al.\ 2009, \textit{ApJ}, 703, 1323

\bibitem[Do et al.(2013)]{2013ApJ...764..154D} Do, T., Lu, J.~R., Ghez, A.~M., et al.\ 2013, \textit{ApJ}, 764, 154 

\bibitem[Ghez et al.(2008)]{2008ApJ...689.1044G} Ghez, A.~M., Salim, S., Weinberg, N.~N., et al.\ 2008, \textit{ApJ}, 689, 1044-1062 

\bibitem[Sch{\"o}del et al.(2010)]{2010A&A...511A..18S} Sch{\"o}del, R., Najarro, F., Muzic, K., \& Eckart, A.\ 2010, \textit{A\&A}, 511, A18

\bibitem[Yelda et al.(2014)]{2014ApJ...783..131Y} Yelda, S., Ghez, A.~M., Lu, J.~R., et al.\ 2014, \textit{ApJ}, 783, 131

\end{thebibliography}
\end{document}